\documentclass[12pt]{article}
\usepackage{epsf} 
\renewcommand{\thefootnote}{\fnsymbol{footnote}} 
 
\def\g{\Gamma}  
\def\gg{{\rm I}\!\g}

\newcommand{\gh}{\hat{\g}}

\begin{document}  

\title{\vskip-3cm{\baselineskip14pt 
\centerline{\normalsize\hfill CERN-TH/2000-304} 
\centerline{\normalsize\hfill DESY 00-157} 
\centerline{\normalsize\hfill NYU-TH/00/01/03} 
\centerline{\normalsize\hfill hep-ph/0011067} 
\centerline{\normalsize\hfill November 2000}} 
\vskip.7cm{\bf  
  Non-invariant two-loop counterterms\\ 
  for the three-gauge-boson vertices} 
\vskip.3cm} 
 
\author{ 
  {P.A. Grassi}$^{a}$, 
  {T. Hurth}$^{b}$, 
  \,and 
  {M. Steinhauser}$^{c}$ 
  \\[3em] 
  {\small \it (a) Physics Department, New York University,}\\ 
  { \small \it  4 Washington Place, New York, NY 10003, USA.} 
  \\[.5em] 
  {\small \it (b) Theory Division, CERN,} \\ 
  {\small \it CH-1211 Geneva 23, Switzerland.} 
  \\[.5em] 
  {\small \it (c) II. Institut f\"ur Theoretische Physik, 
  Universit\"at Hamburg, }\\  
  {\small \it Luruper Chaussee 149, 22761 Hamburg, Germany.}\\ 
} 
\date{} 
\maketitle 
 
\begin{abstract} 
\noindent 
Some practical applications of algebraic renormalization are 
discussed.  In particular we consider the two-loop QCD corrections to 
the three-gauge-boson vertices involving photons, $Z$ and $W$ bosons. 
For this purpose also the corresponding 
two-point functions have to be discussed.  A recently developed 
procedure is used to analyze the breaking terms of the functional 
identities and explicit formulae for the universal counterterms are 
provided. Special attention is devoted to the treatment of infra-red 
divergences. 
\end{abstract}

\thispagestyle{empty} 
\newpage 
\setcounter{page}{1} 
 
 
\renewcommand{\thefootnote}{\arabic{footnote}} 
\setcounter{footnote}{0} 
 
\section{Introduction} 
\label{introduction} 
 
The impressive experimental precision mainly reached at the  
electron--positron  
colliders LEP and SLC and at the proton anti--proton collider 
Tevatron has made it mandatory to evaluate higher order quantum 
corrections.  The dominant contributions arise from perturbative 
calculations in the Standard Model (SM) of elementary particle physics 
and some of its extensions.  As the momentum integrals occurring 
within the usual evaluation of quantum corrections are divergent, a 
regularization accompanied by a renormalization prescription is 
adopted. 
Due to chiral couplings involving $\gamma_5$,  
no invariant 
regularization scheme is known for the Standard 
Model --- leaving aside the lattice  
regularization with the  Ginsparg-Wilson version  of chiral symmetry 
\cite{luescher}.   
The practicality of the latter scheme for higher-loop 
calculations  has to be explored. 
 
It is well known that in the framework of dimensional regularization 
only the non-invariant 't Hooft-Veltman scheme for $\gamma_5$ is shown 
to be consistent to all orders \cite{hoo,martin}. The naive  
dimensional scheme (NDR) 
leads to inconsistencies in connection to $\gamma_5$ and  
the higher order calculations within the SM have already reached a point 
where these inconsistencies cannot be avoided. 
In \cite{jegerlehner} it was emphasized that  
the NDR scheme can still be used in many specific 
calculations and also a practical modification of the  
NDR scheme was proposed.  
In this paper we want to advertise an efficient consistent 
calculation using a  non-invariant regularization scheme.  
This has the consequence that in general the functional 
identities like the Ward-Takahashi (WTI) and the Slavnov-Taylor 
identities (STI) are violated by local breaking terms.  However, the 
concept of algebraic renormalization provides a powerful tool to fix the  
identities and remove the breaking terms (see, e.g., \cite{pige}). 
 
In a recent paper, algebraic renormalization has been considered with 
regard to practical applications~\cite{GraHurSte99}. A procedure has 
been suggested and worked out, which allows an efficient determination 
of the breaking terms.  Actually the computation can be reduced to the 
evaluation of universal, i.e. regularization-scheme-independent, 
counterterms. 
 
In this letter we want to apply the method to the three-gauge-boson 
vertices involving two $W$ bosons and a photon (AWW) or $Z$ boson (ZWW), 
respectively.  They constitute a building block to the important $W$ 
pair production process in $e^+e^-$ annihilation,  
which plays a crucial role at LEP2. 
Furthermore we consider the vertex functions involving three neutral 
gauge bosons, which we will denote by ZAA, AZZ and ZZZ.  
Note that AAA vanishes because of Fury's theorem. 
Also in the context of anomalous couplings the precise study of 
the three-gauge-boson vertices is of importance.

In~\cite{Beenakker} the one-loop diagrams contributing to $e^+e^-\to 
W^+ W^-$ 
have been computed in the framework of dimensional regularization. 
However, proceeding to higher orders, a consistent treatment of 
$\gamma_5$ becomes mandatory and the popular, naive dimensional 
regularization has to be given up. The method of algebraic 
renormalization provides the possibility to adopt any convenient 
regularization --- it only has to be consistent.  
 
Our aim is to focus on two-loop QCD corrections, which has the 
consequence that at the one-loop order only the fermionic contributions  
have to be considered. 
Furthermore we decided to 
work in the framework of the background field gauge, which has the 
advantage that only WTIs with external background 
fields (and no STIs) have to be considered at the highest order.   
They have the same structure to any order in perturbation theory. 
 
Let us in the following briefly review the main steps elaborated 
in~\cite{GraHurSte99} to remove the breaking terms. The use of a 
non-invariant regularization scheme induces breaking terms into the 
STIs 
\begin{equation} 
  \label{STI} 
 \left[ {\cal S}\left( \g \right) \right]^{(n)}  =  
 \hbar^n \Delta_S^{(n)} + {\cal 
    O}(\hbar^{n+1})\,,  
\end{equation} 
which implement the Becchi--Rouet--Stora--Tyutin (BRST) 
symmetry~\cite{brs}, and into the WTIs 
\begin{equation} 
  \label{WTI} 
  {\cal W}_{(\lambda)}\left( \g^{(n)} \right)  = \hbar^n 
  \Delta_W^{(n)}(\lambda) +  
  {\cal O}(\hbar^{n+1})\,,  
\end{equation}  
which implement the background gauge invariance of the SM.  The local 
breaking terms are denoted by $\Delta_S^{(n)}$ and 
$\Delta_W^{(n)}(\lambda)$.  Note that the locality is a consequence of 
the Quantum Action Principle (QAP)~\cite{QAP}.   
Here and in the 
following $\Gamma^{(n)}$ denotes the $n$-loop order, regularized and 
(minimally) subtracted, one-particle-irreducible (1PI) function.  Note 
that the STIs and the WTIs are not able to fix the Green 
functions completely.  Indeed it is possible to add invariant local terms to  
the action, changing the normalization conditions of the functions. A 
complete analysis of the normalization conditions for the SM can be found, 
for instance,  
in~\cite{Sirlin:1980nh,aoki,BoeHolSpi86,grassi,krau_ew}. 
 
The application of a Taylor subtraction of the form 
$(1-T^\delta)$ on Eqs.~(\ref{STI}) and~(\ref{WTI})  
transforms them into 
\begin{eqnarray}\label{WTI_1} 
 \left[ {\cal S}\left( \gh \right) \right]^{(n)}  
 = \hbar^n \Psi_S^{(n)} + {\cal 
    O}(\hbar^{n+1})\, 
  &\mbox{and}& 
  {\cal W}_{(\lambda)}\left( \gh^{(n)} \right)   
  = \hbar^n \Psi_W^{(n)}(\lambda) + {\cal 
    O}(\hbar^{n+1})\,, 
\end{eqnarray} 
where $\hat\Gamma^{(n)} = (1-T^{\delta^\prime})\Gamma^{(n)}$.  
A precise definition 
of $T^\delta$ and $T^{\delta^\prime}$ can be found in  
to~\cite{GraHurSte99}. We only want to mention that 
$\delta$ has to be chosen in such a way that  
$(1-T^\delta)\Delta_{S/W}^{(n)}=0$ and $\delta^\prime$  
corresponds to the power counting degree of the Green functions $\Gamma^{(n)}$. 
The new breaking terms $\Psi_S^{(n)}$ and $\Psi_W^{(n)}(\lambda)$  
are due to over-subtractions and  
can be expressed in terms of a linear combination of ultra-violet (UV) 
finite Green functions and, thus, are independent of the regularization  
scheme~\cite{GraHurSte99}.   
Here we assumed that up to the $(n-1)$-loop order the Green functions are  
already renormalized correctly.  The main difference between  
$\Psi_S^{(n)}$ and $\Psi_W^{(n)}(\lambda)$ is due to the linearity of the 
corresponding operators ${\cal S}$ and ${\cal W}_{(\lambda)}$, respectively.  
In the former case one has to consider non-linear terms arising 
from lower-loop orders. On the contrary in the latter the linearity of the WTI  
simplifies the evaluation of the breaking terms  
and counterterms enormously. 
Finally we introduce 
\begin{eqnarray}   
  \gg^{(n)}  &=& \gh^{(n)} + \Xi^{(n)} = (1-T^{\delta^\prime})  
  \g^{(n)} + \Xi^{(n)} 
  \label{eq:gg} 
  \,, 
\end{eqnarray} 
where the counterterm  
$\Xi^{(n)}$ is chosen in such a way 
that the following identities are fulfilled: 
\begin{eqnarray} 
  \left[{\cal S}\left( \gg \right)\right]^{(n)} = 0\,, 
  && 
  {\cal W}_{(\lambda)}\left( \gg^{(n)} \right) =0\,. 
  \label{eq:conds} 
\end{eqnarray} 
In general it is quite simple to compute the total counterterm 
$(- T^{\delta^\prime} \g^{(n)} + \Xi^{(n)})$, as it can be expressed in  
terms of Green functions expanded around zero external momenta. 
 
As already mentioned above, there is still the freedom  
to add invariant counterterms, $\Xi_N^{(n)}$, to $\gg^{(n)}$ 
in Eq.~(\ref{eq:gg}). 
In other words, we have the freedom to  
impose normalization conditions, which lead, in addition to 
Eqs.~(\ref{eq:conds}), to the equations 
\begin{eqnarray} 
  {\cal N}_i \left( \gg^{(n)} \right) = 0\,, 
  \label{nor_con} 
\end{eqnarray} 
where the index $i$ runs over all independent parameters of the SM.   
As the Green function $\Gamma^{(n)}$ also has to fulfill this condition,  
we have for the counterterm 
\begin{eqnarray}\label{nc} 
  {\cal N}_i \left( -T^\delta\g^{(n)} + \Xi^{(n)} + \Xi_N^{(n)} 
  \right)   
  &=& 0\,, 
\end{eqnarray} 
which is a local equation. This means that, whenever the effort to 
impose the normalization conditions is made, the changes due to the 
subtraction are only local changes, which can be easily compensated. 
Explicit examples will be discussed at the end of Section~\ref{one}. 
Notice that the imposition of 
normalization conditions is a very important ingredient of the 
computation in order to compare with other schemes and in order to 
simplify the breaking terms themselves.

The procedure described so far is based on the Taylor operator 
$T^\delta$. In the presence of massless particles, this may introduce 
IR divergences. In the examples discussed in this paper, 
eventual IR problems are encountered in intermediate steps after 
neglecting one of the quark masses, a well-justified approximation in 
the case of the top--bottom doublet. 
Note that this kind of IR divergences  
should not be confused with those arising in connection with on-shell 
conditions of internal particle propagators. 
The appropriate methods for dealing with IR divergences are introduced 
in Section~\ref{IRR}. 
 
Although we are mainly interested in the three-point functions, also 
some two-point functions with external (background) gauge bosons have 
to be treated properly in order to be able to renormalize 
the amplitudes correctly. They will be discussed in Section~\ref{sec:2point}. 
 
In Section~\ref{one} the 
one-loop sub-diagrams occurring in the two-loop calculation are 
analyzed. 
In Section~\ref{sec:neutral} the vertices 
involving only neutral gauge bosons are considered and, after 
introducing the two-point functions, we are ready to discuss the  
cases $AWW$ and $ZWW$ in Section~\ref{sec:charged}. 
 
 
\section{\label{one}One-loop Green functions} 
 
This section is devoted to the one-loop sub-diagrams  
induced by QCD, which are needed for the renormalization. 
 
In the case of neutral gauge bosons one has to take into account 
the two- and three-point Green functions\footnote{ 
All momenta are considered as incoming. In the Green functions 
      $\Gamma_{\phi_1 \dots \phi_n}$ they are assigned to the  
      corresponding fields starting from the right. 
      The momentum of the most left field  
      is determined via momentum conservation.} 
$\g^{(1)}_{\bar q q}(p)$ and 
$\g^{(1)}_{\hat A_\mu\bar q q}(p,q)$ 
and the corresponding vertices where the photon is replaced by the $Z$ 
boson. For $q$ we have $q\in\{u,d\}$, where $u$ and $d$  
refer to a generic quark doublet.  
After the analysis of the WTIs, also the 
vertices with the neutral Goldstone boson, $\hat G^0$, 
$\g^{(1)}_{\hat G^0\bar u u}(p,q)$ 
and 
$\g^{(1)}_{\hat G^0\bar d d}(p,q)$, 
turn out to be relevant.  
For the amplitudes AWW and ZWW the vertices  
$\g^{(1)}_{\hat W^+_\mu\bar u d}(p,q)$  
and  
$\g^{(1)}_{\hat G^+ \bar u d}(p,q)$  
are needed in addition. 
As we work in the framework of the BFM, no Green functions with external 
scalar or gauge fields have to be considered, and we are left with only 
three WTIs: 
\begin{eqnarray} 
  \label{one_1} 
i \, (p+ q)^\mu   \g^{(1)}_{\hat A_\mu\bar q q}(p,q) + i  e   
\, Q_q\left[   \g^{(1)}_{\bar q q}(q) - \g^{(1)}_{\bar q q}(-p) 
\right]  
  &=&  
\Delta^{(1)}_{W,\lambda_A \bar q q}(p,q) 
\,,  
\nonumber \\ 
i \, (p+ q)^\mu   \g^{(1)}_{\hat Z_\mu\bar q q}(p,q)  - M_Z  
\g^{(1)}_{\hat G^0 \bar q q}(p,q)  
\hphantom{xxxxxxxxxxxxxxxxxx} 
\nonumber \\ 
- i e \left[  ( v_q - a_q \gamma_5) 
  \g^{(1)}_{\bar q q}(q) -  
\g^{(1)}_{\bar q q}(-p) (v_q + a_q \gamma_5)  \right] &=&  
\Delta^{(1)}_{W,\lambda_Z \bar q q}(p,q)  
\,, 
\nonumber\\  
i \left(p+q\right)_\nu \g^{(1)}_{\hat W^{+}_{\nu} \bar{u} 
    d}(p,q)      
  +  i\, M_W  \g^{(1)}_{\hat G^{+} \bar{u} d}(p, q)  
\hphantom{xxxxxxxxxx} 
\nonumber   \\ \mbox{} 
 + { i e \over s_W\sqrt{2}} \left[ 
    \g^{(1)}_{\bar{u} u^\prime}(- p)  
    V_{u' d} P_L 
    -  V_{{u} d^\prime} P_R 
    \g^{(1)}_{\bar{d}' d}(q)   \right] &=&   
  \Delta^{(1)}_{W,\lambda_+  \bar{u} d}(p,q) 
\,. 
\end{eqnarray} 
Here and in the following we define the Weinberg angle through 
$c_W=M_W/M_Z$ as we want to maintain the form of the WTIs to be the 
same to all orders. 
The couplings of the fermions to the $Z$ boson are given by 
$v_q=(I^3_{q} - 2 s^2_W Q_q)/ (2 c_W s_W)$ and  
$a_q =I^3_{q}/(2c_W s_W)$, where $I^3_{q}$ and $Q_q$ are the third 
generator of $SU_W(2)$ and the electric charge of the $q$ quark, 
respectively. 
The equation for $\Delta_{W,\lambda_-\bar{d}u}$ has been omitted as it 
can easily  
be obtained from the last one in~(\ref{one_1}). 
$V_{q q^\prime}$ are the CKM matrix elements where the summation over 
the primed quantities is understood and 
$P_{L / R}=(1\mp\gamma_5)/2$ are the chiral projectors.  
 
In order to remove the breaking terms we apply the  
Taylor operator $(1-T^1_{p,q})$ as the $\Delta$'s 
have mass dimension one. This leads to 
\begin{eqnarray} 
  \label{one_1.1} 
  i(p+q)^\mu   \gh^{(1)}_{\hat A_\mu\bar q q}(p,q) + i e Q_q  
    \left[ \gh^{(1)}_{\bar q q}(q) -  
    \gh^{(1)}_{\bar q q}(-p) \right]  
  &=& 0\,,  
  \nonumber\\ 
  i(p+ q)^\mu \gh^{(1)}_{\hat Z_\mu\bar q q}(p,q)  
  -  M_Z \gh^{(1)}_{\hat G^0 \bar q q}(p,q)  
  \hphantom{xxxxxxxxxxxxxxxxxx} 
  \nonumber \\ 
- i e \left[  ( v_q - a_q \gamma_5) 
  \gh^{(1)}_{\bar q q}(q) -  
\gh^{(1)}_{\bar q q}(-p)  (v_q + a_q \gamma_5)  \right] &=&  
 \Psi^{(1)}_{W,\lambda_Z \bar q q}(p,q) \,, 
  \nonumber \\ 
  i \left(p + q \right)_\nu \gh^{(1)}_{\hat W^{+}_{\nu} \bar{u} 
    d}(p , q)      
  +  i\, M_W  \gh^{(1)}_{\hat G^{+} \bar{u} d}(p , q)  
  \hphantom{xxxxxxxx} 
  \nonumber \\ 
  + { i e \over s_W\sqrt{2}} \left[ 
    \gh^{(1)}_{\bar{u} u^\prime}(- p)  
   V_{u' d} P_L 
   -  V_{{u} d^\prime} P_R 
   \gh^{(1)}_{\bar{d}^\prime d}(q)  
  \right]  
  &=&   
  \Psi^{(1)}_{W,\lambda_+  \bar{u} d}(p,q) 
  \,. 
\end{eqnarray} 
The $\Psi$'s, which occur by over-subtraction, are finite and read: 
\begin{eqnarray} 
  \Psi^{(1)}_{W,\lambda_Z \bar q q}(p,q)  
  &=&  
 -  M_Z \left(p^\rho \partial_{p^\rho } + q^\rho \partial_{q^\rho }\right)   
  \g^{(1)}_{\hat G^0 \bar q q}(p,q) \Bigg|_{p=q=0}  
  \nonumber\\ 
  &=&  
  i {\alpha_s \over 4\pi} C_F {e \over s_W c_W} I^3_q  
  \left(\frac{1}{2}\xi-2\right)  
  (\not\!p\, + \!\not\!q) \gamma_5  
   \,, 
  \nonumber\\  
  \Psi^{(1)}_{W,\lambda_+  \bar{u} d}(p,q)  &=& 
  i\, M_W  \left( p^\rho \partial_{p^\rho} + q^\rho 
    \partial_{q^\rho} \right)   
  \g^{(1)}_{\hat G^{+} \bar{u} d}(p,q) \Bigg|_{p=q=0} 
  \nonumber\\ 
  &=& 
  i {\alpha_s \over 4\pi} C_F {e V_{ud}\over 2\sqrt{2}s_W}  
  \left[ 
  \left(1+\frac{1}{2}\xi\right) 
  (\not\!p\, + \!\not\!q) P_L 
  \right.\nonumber\\&&\left.\mbox{} 
  + \frac{\left(1-\frac{1}{2}\xi\right)\ln\frac{m_u^2}{m_d^2}}{m_u^2-m_d^2} 
  \left( 
    \left( m_u^2 \not\!p\, + m_d^2 \!\not\!q \right) P_L 
   -2 m_u m_d (\not\!p\, + \!\not\!q) P_R 
  \right) 
  \right] 
  \,, 
  \label{one_2} 
\end{eqnarray} 
where the gauge parameter $\xi$ is defined through the gluon 
propagator 
$D_g(q)=i(-g^{\mu\nu}+\xi q^\mu q^\nu/q^2)/(q^2+i\epsilon)$ 
and $C_F=4/3$ is the colour factor. 
Owing to the linear momentum dependence of  
$\Psi^{(1)}_{W,\lambda_Z \bar q q}(p,q)$ and  
$\Psi^{(1)}_{W,\lambda_+ \bar{u} d}(p,q)$, 
their contribution can be absorbed by the counterterms 
\begin{eqnarray} 
  \label{one_3} 
  \Xi^{(1),W}_{\hat Z_\mu \bar q q}(p,q)  
  &=&  \xi^{(1)}_{V \bar q q} \gamma^\mu +  
  \xi^{(1)}_{A\bar q q} \gamma^\mu \gamma_5   
  \,,  
  \nonumber \\ 
  \Xi^{(1),W}_{\hat W^{+}_{\mu} \bar{u} d}(p,q)  
  &=&  \xi^{(1)}_{L, \hat W^{+} \bar{u} d} \, \gamma^\mu   P_L\,   
       + \xi^{(1)}_{R, \hat W^{+} \bar{u} d} \, \gamma^\mu   P_R  
 \,.  
\end{eqnarray} 
for three point functions and  
\begin{eqnarray}\label{one_5} 
  \Xi^{(1),N}_{\bar q q}(p) &=&  
  \xi^{(1)}_{2,q} \left(\not\!p - m_q \right) + 
  \xi^{(1)}_{q} m_q \,, 
\end{eqnarray} 
for quark self-energies.  
The coefficients $\xi^{(1)}_{2,q}$ and 
$\xi^{(1)}_{q}$ have to be tuned for the two-point function 
\begin{eqnarray} 
  \label{nor_con_q} 
  \gg^{(1)}_{\bar{q} q}(p) &=& \g^{(1)}_{\bar{q} q}(p)  - \left[T^1_p 
  \g^{(1)}_{\bar{q} q}(p)\right]  +   \Xi^{(1),N}_{\bar q q}(p) 
  \,, 
\end{eqnarray} 
to restore the WTI (\ref{one_1.1}) and to satisfy the specific normalization conditions. 
In the case of the on-shell scheme, the condition 
$\gg^{(1)}_{\bar{q} q}(p^*) =0$, for instance, where the real part of $p^*$  
corresponds to the physical quark mass, would fix $\xi_q^{(1)}$. 
 
From the explicit results in Eq.~(\ref{one_2}) one can see that the 
vector coefficient of $\Xi^{(1),W}_{\hat Z_\mu \bar q q}$ 
has to be zero. Concerning the axial-vector 
part, there are in principle two structures. However, since   
only the combination $(\not\!p\, + \!\not\!q)$ 
appears in Eq.~(\ref{one_2}) for 
$\Psi^{(1)}_{W,\lambda_Z \bar q q}(p,q)$, 
$\xi^{(1)}_{A\bar{q}q}$ is sufficient to remove 
the breaking term and we thus have 
\begin{eqnarray} 
  \label{one_3.1} 
  \xi^{(1)}_{V\bar q q}= 0\,,&& \xi^{(1)}_{A \bar q q}=  
  {\alpha_s \over 4\pi} C_F {e \over s_W c_W} I^3_q  
  \left(\frac{1}{2}\xi-2\right)  
   \,.  
\end{eqnarray} 
Similarly we get in the case of  
$\Xi^{(1),W}_{\hat W^{+}_{\mu} \bar{u} d}(p,q)$ 
\begin{eqnarray} 
  \label{one_3.2} 
 && 
        \xi^{(1)}_{2,u} = \xi^{(1)}_{2,d} -    
        {\alpha_s \over 8\pi} C_F  \left(1-\frac{1}{2}\xi\right) 
 \ln\left( \frac{m^2_u}{m^2_d} \right)\,, \nonumber \\  
 && \xi^{(1)}_{L, \hat W^{+} \bar{u} d}= {e  V_{ud}\over 2 \sqrt{2}s_W} 
 \left( \xi^{(1)}_{2,u} - \xi^{(1)}_{2,d} \right) +   
  {\alpha_s \over 4\pi} C_F {e V_{ud}\over 2\sqrt{2}s_W}  \left[ 
 \left(1+\frac{1}{2}\xi\right) +  
\frac{\left(1-\frac{1}{2}\xi\right) m^2_u 
 \ln\frac{m_u^2}{m_d^2}}{m_u^2-m_d^2} \right]\,, \nonumber \\  
 && \xi^{(1)}_{R, \hat W^{+} \bar{u} d}=- 
  {\alpha_s \over 4\pi} C_F {e V_{ud}\over \sqrt{2}s_W} \frac{  m_u m_d 
 \left(1-\frac{1}{2}\xi\right)\ln\frac{m_u^2}{m_d^2}}{m_u^2-m_d^2}  
\,.  
\end{eqnarray} 
The free parameter $ \xi^{(1)}_{2,d}$ can be fixed by a normalization condition on the  
two-point function residue.  
 
Finally we can write down the symmetric one-loop Green functions 
for the neutral and charged current vertices: 
\begin{eqnarray} 
  \gg^{(1)}_{\hat A_\mu\bar q q}(p,q)  
  &=& 
  \g^{(1)}_{\hat A_\mu\bar q q}(p,q)  - \left[ T^0_{p,q} 
  \g^{(1)}_{\hat A_\mu\bar q q}(p,q) \right] -  
  e Q_q  \xi^{(1)}_{2,q}  \gamma^\mu  
  \nonumber \,, \\ 
  \gg^{(1)}_{\hat Z_\mu\bar q q}(p,q)  
  &=&  
  \g^{(1)}_{\hat Z_\mu\bar q q}(p,q) - \left[  T^0_{p,q} 
  \g^{(1)}_{\hat Z_\mu\bar q q}(p,q)   
  - \xi^{(1)}_{A \bar q q} 
  \gamma^\mu \gamma_5 
  \right]  
  \nonumber\\&&\mbox{} 
  + e \xi^{(1)}_{2,q} \gamma^\mu \Big( 
  v_q  - a_q\gamma_5 \Big) \,,   
  \nonumber\\   
  \gg^{(1)}_{\hat G^0 \bar q q}(p,q)  
  &=&  
  \g^{(1)}_{\hat G^0 \bar q q}(p,q) - \left[ T^0_{p,q} \g^{(1)}_{\hat 
  G^0 \bar q q}(p,q) \right] -   
   { e I^3_{q} m_q \over  i s_W M_W} \left( \xi^{(1)}_q  - 
  \xi^{(1)}_{2,q} \right)  
  \gamma^5  
  \,, 
  \nonumber\\ 
  \gg^{(1)}_{\hat W^{+}_{\mu} \bar{u} d}(p,q)   
  &=&  
  \g^{(1)}_{\hat W^{+}_{\mu} \bar{u} d}(p,q) 
  - \left[ T^0_{p,q} \g^{(1)}_{\hat W^{+}_{\mu} \bar{u} d}(p,q)   
  - \xi^{(1)}_{L, \hat W^{+}\bar{u} d} \gamma^\mu P_L 
  - \xi^{(1)}_{R, \hat W^{+}\bar{u} d} \gamma^\mu P_R \right]  
  \nonumber\\&&\mbox{} 
  +  {e V_{ud} \over 2 \sqrt{2} s_W} \left(\xi^{(1)}_{2,u} + \xi^{(1)}_{2,d} 
  \right) \gamma^\mu P_L  
  \,,  
  \nonumber \\ 
  \gg^{(1)}_{\hat G^{+} \bar{u} d}(p,q)  
  &=& 
  \g^{(1)}_{\hat G^{+} 
  \bar{u} d}(p,q)   
  - \left[ T^0_{p,q} \g^{(1)}_{\hat G^{+} \bar{u} d}(p,q) \right]   
  \nonumber \\ 
  &&- 
  { e V_{ud} m_u \over \sqrt{2} s_W M_W}\left( \xi^{(1)}_u  - 
  \xi^{(1)}_{2,u} \right) P_L +  
   { e V_{ud} m_d \over \sqrt{2} s_W M_W} \left( \xi^{(1)}_d - 
  \xi^{(1)}_{2,d} \right) P_R   
  \,. 
  \label{one_6} 
\end{eqnarray} 
In this specific sector only the counterterms $\xi^{(1)}_u, \xi^{(1)}_d$ and  
$\xi^{(1)}_{2,d}$ can be tuned to imposed suitable normalization conditons, the others  
are indeed necessary to restore the WTI~(\ref{one_1.1}).

Using the notation of the introduction, Eqs.~(\ref{one_6})  
                 can be expressed in the following compact form 
\begin{eqnarray}   
  \gg^{(n)}  &=& \gh^{(n)} + \Xi^{(n)} 
  \nonumber\\ 
  &=& \g^{(n)} - \left[ T^\delta 
  \g^{(n)} + \Xi^{(n)}_W \right]  +   
  \Xi^{(n)}_N  
  \nonumber \\ 
  &=& 
  \g^{(n)}_{\rm bare} - \g^{(n)}_{\rm UV} -  \left[ T^\delta 
  \g^{(n)}_{\rm bare}  -   
  T^\delta \g^{(n)}_{\rm UV} + \Xi^{(n)}_W \right]  +  \Xi^{(n)}_N \,, 
  \label{eq:gg_1} 
\end{eqnarray} 
In the third line we have introduced the bare Green function  
$\g_{\rm bare}^{(n)}$. 
This quantity is defined by $\g^{(n)} = 
 \g^{(n)}_{\rm bare} - \g^{(n)}_{\rm UV}$, where  $\g^{(n)}_{\rm UV}$  
denotes the necessary UV counterterms computed in the specified  
regularization.  
Clearly, the complete one-loop counterterms, namely  
 $\gg^{(n)} - \g^{(n)}_{\rm bare}$, have to be taken into account  
at the two-loop level.

 
\section{\label{sec:neutral}Neutral-gauge-boson vertices} 
 
\subsection{\label{AZZ}The AZZ case} 
 
The vertex involving a photon and two $Z$ bosons is used to 
demonstrate the main features of our technique. This 
example clarifies also the issue of anomaly cancellation in our formalism.  
In principle there is also the vertex ZAA. However, it is very 
similar to AZZ. Thus we will not present explicit results for ZAA as 
the corresponding equations are simply obtained by replacing 
one of the $Z$ bosons by a photon. 
 
As we are dealing with external background fields, WTIs can be used to fix 
the counter\-terms for these amplitudes.  In order to derive the 
complete set of identities, one of the gauge fields has to be replaced 
by the infinitesimal parameter of the background gauge transformations 
and then the derivatives of the functional WTI have to be performed (cf. 
Ref.~\cite{GraHurSte99}).  This leads to six identities, which 
naturally split into two sets depending on whether the index of the 
photon or the $Z$ boson is contracted with the external momentum. 
We get the following closed (under renormalization) set of equations 
\begin{eqnarray} 
   \label{azz_1} 
   i (p+q)^\mu \g^{(n)}_{\hat{A}_\mu \hat{Z}_\nu \hat{Z}_\rho}(p,q)  
   &=&  
    \Delta^{(n)}_{W, \lambda_{A} \hat{Z}_\nu \hat{Z}_\rho}(p,q)  \,, 
\nonumber\\ 
    \label{azz_2} 
   - i p^\nu \g^{(n)}_{\hat{A}_\mu \hat{Z}_\nu \hat{Z}_\rho}(p,q) - M_{Z}  
   \g^{(n)}_{\hat{A}_\mu \hat{G}^0 \hat{Z}_\rho}(p,q)  
   &=& 
   \Delta^{(n)}_{W, \hat{A}_\mu  \lambda_Z \hat{Z}_\rho}(p,q)  \,, 
\nonumber\\ 
   \label{azz_3} 
   i (p+q)^\mu \g^{(n)}_{\hat{A}_\mu \hat{G}^0 \hat{Z}_\rho}(p,q)  
   &=& 
   \Delta^{(n)}_{W, \lambda_A  \hat{G}^0 \hat{Z}_\rho}(p,q)  \,,  
\nonumber\\ 
   \label{azz_4} 
   - i p^\nu \g^{(n)}_{\hat{A}_\mu \hat{Z}_\nu \hat{G}^0}(p,q) - M_Z 
   \g^{(n)}_{\hat{A}_\mu \hat{G}^0 \hat{G}^0}(p,q)   
   &=& \Delta^{(n)}_{W, \hat{A}_\mu  \lambda_Z \hat{G}^0}(p,q)  \,, 
\nonumber\\ 
   \label{azz_5} 
   i (p+q)^\mu \g^{(n)}_{\hat{A}_\mu \hat{G}^0 \hat{G}^0}(p,q)  
   &=& 
   \Delta^{(n)}_{W, \lambda_A  \hat{G}^0 \hat{G}^0}(p,q)  \,.  
\end{eqnarray} 
There are in principle three more equations where the contraction is 
performed with $q^\rho$. However, they contain no new information. 
 
The breaking terms in Eqs.~(\ref{azz_5}) have mass 
dimension two. Thus we have to apply the operator 
$(1-T^{2}_{p,q})$ in order to remove them. This leads to 
\begin{eqnarray} 
   \label{azz_1.1} 
    i (p+q)^\mu \gh^{(n)}_{\hat{A}_\mu \hat{Z}_\nu \hat{Z}_\rho}(p,q)  
    &=&  
    \Psi^{(n)}_{W, \lambda_A \hat{Z}_\nu \hat{Z}_\rho}(p,q)  \,\,=\,\, 0 \,, 
\\ 
    \label{azz_2.1} 
    -i p^\nu \gh^{(n)}_{\hat{A}_\mu \hat{Z}_\nu \hat{Z}_\rho}(p,q) - 
    M_Z  \gh^{(n)}_{\hat{A}_\mu \hat{G}^0 \hat{Z}_\rho}(p,q)   
    &=& \Psi^{(n)}_{W, \hat{A}_\mu  \lambda_Z \hat{Z}_\rho}(p,q)  \,, 
\\ 
    \label{azz_3.1} 
    i (p+q)^\mu \gh^{(n)}_{\hat{A}_\mu \hat{G}^0 \hat{Z}_\rho}(p,q)  
    &=&  
    \Psi^{(n)}_{W, \lambda_A  \hat{G}^0 \hat{Z}_\rho}(p,q)  \,\,=\,\, 0 \,,  
\\ 
    \label{azz_4.1} 
    -i p^\nu \gh^{(n)}_{\hat{A}_\mu \hat{Z}_\nu \hat{G}^0}(p,q) - M_Z 
    \gh^{(n)}_{\hat{A}_\mu \hat{G}^0 \hat{G}^0}(p,q)   
    &=& 
    \Psi^{(n)}_{W, \hat{A}_\mu  \lambda_Z \hat{G}^0}(p,q) \,\,=\,\, 0 \,, 
\\ 
    \label{azz_5.1} 
    i (p+q)^\mu \gh^{(n)}_{\hat{A}_\mu \hat{G}^0 \hat{G}^0}(p,q)  
    &=& 
    \Psi^{(n)}_{W, \lambda_A  \hat{G}^0 \hat{G}^0}(p,q)  \,\,=\,\, 0 \,,  
    \end{eqnarray} 
where only $\Psi^{(n)}_{W,\hat{A}_\mu  \lambda_Z \hat{Z}_\rho}(p,q)$ 
is non-vanishing: 
\begin{eqnarray} 
  \label{azz_psi} 
  \Psi^{(n)}_{W, \hat{A}_\mu  \lambda_Z \hat{Z}_\rho}(p,q)  
  &=&  
  - {M_Z \over 2}  
  \left(  
  p^\alpha p^\beta   \partial_{p^\alpha} \partial_{p^\beta} + 2 
  p^\alpha q^\beta   \partial_{p^\alpha} \partial_{q^\beta} +  
  q^\alpha q^\beta   \partial_{q^\alpha} \partial_{q^\beta}   
  \right) \gh^{(n)}_{\hat{A}_\mu \hat{G}^0 \hat{Z}_\rho}(p,q) 
  \Bigg|_{p=q=0} 
  \,. 
  \nonumber\\ 
\end{eqnarray} 
The other breaking terms are zero because   
of QED-like WTIs for the external background photon  
(Eqs.~(\ref{azz_1.1}),~(\ref{azz_3.1}) and~(\ref{azz_5.1})) 
and Lorentz invariance (Eq.~(\ref{azz_4.1})), respectively. 
 
In order to remove  
$\Psi^{(n)}_{W,\hat{A}_\mu  \lambda_Z \hat{Z}_\rho}(p,q)$, 
a counterterm, 
$\Xi^{(n)}_{\hat{A}_\mu \hat{Z}_\nu \hat{Z}_\rho}$, has to be introduced 
for the Green functions 
$\gh^{(n)}_{\hat{A}_\mu \hat{Z}_\nu \hat{Z}_\rho}(p,q)$.  
Notice, however, that this Green function also 
appears in Eq.~(\ref{azz_1.1}). In order not to spoil 
Eq.~(\ref{azz_1.1}), $\Xi^{(n)}_{\hat{A}_\mu \hat{Z}_\nu \hat{Z}_\rho}$ 
has to be longitudinal w.r.t. the photon index $\mu$. 
On the other hand, if we 
contract Eq.~(\ref{azz_2.1}) by $(p+q)^\mu$ and use 
Eqs.~(\ref{azz_1.1}) 
and~(\ref{azz_3.1}), 
we obtain 
\begin{eqnarray} 
  \label{con_3} 
  (p+q)^\mu  \Psi^{(n)}_{W, \hat{A}_\mu  \lambda_Z 
  \hat{Z}_\rho}(p,q)  
  &=& 0 \,.   
\end{eqnarray} 
This implies that the breaking term  
$\Psi^{(n)}_{W, \hat{A}_\mu \lambda_Z \hat{Z}_\rho}(p,q)$   
should be transversal w.r.t. the photon index $\mu$. 
Thus, combining the two arguments, we deduce that 
the breaking term itself has to be zero. 
 
We have checked this prediction by explicit calculations at the one- and 
two-loop levels. 
At one-loop order, the contribution from one fermion species gives 
\begin{eqnarray} 
  \Psi^{(1)}_{W, \hat{A}_\mu  \lambda_Z \hat{Z}_\rho}(p,q) &=&  
  i\frac{\alpha}{4\pi}\frac{4e}{s_Wc_W}I_3^q Q_q v_q 
  \epsilon^{\mu\rho\alpha\beta} p_\alpha q_\beta 
  \,, 
  \label{eq:psiazz1l} 
\end{eqnarray} 
with $\epsilon^{0123}=1$. 
The only reminder on the fermion type 
is the third component of the isospin, the 
charge and the coupling to the $Z$ boson. 
Thus, after summing over a complete family of quarks and leptons  
one gets zero. This is the same mechanism which leads to the 
cancellation of the Adler--Bardeen anomaly in the SM~\cite{ABJ}. 
At two loops already the sum over all contributing diagrams of one 
quark flavour is zero as we checked by an explicit calculation. 
 
This example provides a nice demonstration of the power of our 
technique. Regardless of the regularization adopted to compute the Green 
functions $\g^{(n)}$, the zero-momentum subtraction fixes 
automatically the non-invariant counterterms needed to restore the 
symmetries. In particular, we found that besides one-loop counterterms 
(which were discussed in Section~\ref{one}) 
no other counterterm is necessary to  
define the properly renormalized amplitudes.  
Finally, the direct computation of  
breaking term  
$\Psi^{(n)}_{W, \hat{A}_\mu  \lambda_Z \hat{Z}_\rho}(p,q)$ 
at one- and two-loop level ($n=1,2$)  
shows how the anomaly coefficient can be computed and the 
Adler--Bardeen non-renormalization theorem   
can be verified in the present framework.  
 
 
\subsection{The ZZZ case } 
\label{ZZZ} 
 
Also in the ZZZ case there is no tree-level contribution, which makes 
it similar to the previous case. Here the closed system of WTIs looks 
as follows 
\begin{eqnarray} 
   \label{zzz_1} 
   i (p+q)^\mu \g^{(n)}_{\hat{Z}_\mu \hat{Z}_\nu \hat{Z}_\rho}(p,q) - M_{Z}  
    \g^{(n)}_{\hat{G}^0  \hat{Z}_\nu \hat{Z}_\rho}(p,q)  
    &=& \Delta^{(n)}_{W, \lambda_Z \hat{Z}_\nu  \hat{Z}_\rho}(p,q)  \,, 
\nonumber\\ 
   \label{zzz_2} 
   i (p+q)^\mu \g^{(n)}_{\hat{Z}_\mu \hat{G}^0 \hat{Z}_\rho}(p,q) - M_{Z}  
    \g^{(n)}_{\hat{G}^0  \hat{G}^0 \hat{Z}_\rho}(p,q)  
    &=& \Delta^{(n)}_{W, \lambda_Z \hat{G}^0  \hat{Z}_\rho}(p,q)  \,, 
\nonumber\\ 
   \label{zzz_3} 
   i (p+q)^\mu \g^{(n)}_{\hat{Z}_\mu \hat{G}^0 \hat{G}}(p,q) - M_{Z}  
    \g^{(n)}_{\hat{G}^0  \hat{G}^0 \hat{G}^0}(p,q)  
    &=& \Delta^{(n)}_{W, \lambda_Z \hat{G}^0  \hat{G}^0}(p,q)  \,. 
\end{eqnarray} 
Taylor subtraction, by application of the operator $(1-T^{2}_{p,q})$, 
partially eliminates the breaking terms and leads to 
\begin{eqnarray} 
   \label{zzz_1.1} 
   \Psi^{(n)}_{W, \lambda_Z \hat{Z}_\nu  \hat{Z}_\rho}(p,q)  
   &=& - {M_Z \over 2}  
   \left[ \left(  
     p^\alpha p^\beta   \partial_{p^\alpha} \partial_{p^\beta} + 2 
     p^\alpha q^\beta   \partial_{p^\alpha} \partial_{q^\beta} +  
     q^\alpha q^\beta   \partial_{q^\alpha} \partial_{q^\beta}   
   \right) \gh^{(n)}_{\hat{G}^0 \hat{Z}_\nu \hat{Z}_\rho}(p,q) 
   \right]_{p=q=0} 
   \,,  
\nonumber \\ 
   \label{zzz_2.1} 
   \Psi^{(n)}_{W, \lambda_Z \hat{G}^0  \hat{Z}_\rho}(p,q) &=& 0 \,, 
\nonumber\\ 
    \label{zzz_3.1} 
   \Psi^{(n)}_{W, \lambda_Z \hat{G}^0  \hat{G}^0}(p,q)  &=& 
   - {M_Z \over 2}  
   \left[ \left(  
     p^\alpha p^\beta   \partial_{p^\alpha} \partial_{p^\beta} + 2 
     p^\alpha q^\beta   \partial_{p^\alpha} \partial_{q^\beta} +  
     q^\alpha q^\beta   \partial_{q^\alpha} \partial_{q^\beta}   
   \right) \gh^{(n)}_{\hat{G}^0 \hat{G}^0 \hat{G}^0}(p,q) \right]_{p=q=0}   
   \,. 
\nonumber\\ 
\end{eqnarray} 
Owing to CP invariance and the fact that we consider only QCD 
corrections, it turns out that both 
$\Psi^{(n)}_{W, \lambda_Z \hat{Z}_\nu  \hat{Z}_\rho}(p,q)$ 
and 
$\Psi^{(n)}_{W, \lambda_Z \hat{G}^0 \hat{G}^0}(p,q)$ are zero at the 
one- and two-loop order, i.e. for $n=1,2$. 
This has been checked by explicit one- and two-loop calculations. 
Actually  
$\Psi^{(n)}_{W, \lambda_Z \hat{Z}_\nu  \hat{Z}_\rho}(p,q)$ shows the 
same behaviour as the corresponding contribution in the AZZ case: for 
$n=1$ the breaking term vanishes after summing over a whole family and  
at two-loop order the contribution from each fermion species gives 
separately zero.  
 
At higher orders, 
CP violation induced by the CKM quark mixing matrix might render   
$\Psi^{(n)}_{W, \lambda_Z \hat{Z}_\nu  \hat{Z}_\rho}(p,q)$ 
and 
$\Psi^{(n)}_{W, \lambda_Z \hat{G}^0  \hat{G}^0}(p,q)$  
different from zero. 
The corresponding counterterm for  
$\g^{(n)}_{\hat Z_\mu \hat Z_\nu \hat Z_\rho}$, 
which has to be introduced in order to remove 
$\Psi^{(n)}_{W, \lambda_Z \hat{Z}_\nu  \hat{Z}_\rho}(p,q)$, 
would read  
\begin{eqnarray} 
  \label{cou_1.1} 
  \Xi^{(n),W}_{\hat Z_\mu \hat Z_\nu \hat Z_\rho}(p,q) &=&  
  \xi^{(n)}_{\hat{Z}\hat{Z}\hat{Z}}   
  \, \left( p^\nu g^{\mu\rho}  + q^\rho 
  g^{\mu\nu} -  (p+q)^\mu g^{\nu\rho} \right)   
  \,, 
\end{eqnarray} 
where the coefficient $\xi^{(n)}_{\hat{Z}\hat{Z}\hat{Z}}$ 
is determined in terms of the breaking terms (\ref{zzz_1.1}).  
Notice that, in the present case, there is no   
condition like Eq.~(\ref{con_3}) and thus, unlike the AZZ case, 
in general there are counterterms of the form~(\ref{cou_1.1}). 
 
 
\section{Fermionic contribution to the two-point functions} 
\label{sec:2point} 
 
A proper renormalization of the two-point functions is needed in 
order to be able to   
correctly renormalize the three-point Green functions AWW and ZWW. 
Moreover, most of the normalization conditions are expressed in 
terms of two-point functions. 
In this section we mainly concentrate on the results needed in the 
forthcoming parts of the paper, while details can be found in   
Ref.~\cite{GraHurSte99}. 
 
Applying our prescriptions to the two-point Green functions with  
external background fields shows that only the self-energies of the 
(background) $W$ and $Z$ bosons are affected by breaking terms. The 
corresponding symmetrical Green functions read 
\begin{eqnarray} 
  \label{nor_con_W_fin} 
  \gg^{(n)}_{\hat W^+_\mu \hat W^-_\nu}(p) &=&  \g^{(n)}_{\hat W^+_\mu 
  \hat W^-_\nu}(p)  
  - \left[ T^2_{p}   \g^{(n)}_{\hat W^+_\mu \hat W^-_\nu}(p)  + 
    \xi^{(n)}_{\hat W,1} \, p^2 g_{\mu\nu}   
+  \xi^{(n)}_{\hat W,2} \, p_\mu p_\nu \right] + \xi^{(n)}_{M_W} g_{\mu\nu} 
  \,, \nonumber \\  
  \gg^{(n)}_{\hat Z_\mu \hat Z_\nu}(p) &=&  \g^{(n)}_{\hat Z_\mu \hat 
  Z_\nu}(p)  
  - \left[T^2_{p}    \g^{(n)}_{\hat Z_\mu \hat Z_\nu}(p)  + 
  \xi^{(n)}_{\hat Z, 1} \, p^2 g_{\mu\nu}  
+  \xi^{(n)}_{\hat Z, 2} \, p_\mu p_\nu \right] + \xi^{(n)}_{M_Z} g_{\mu\nu} 
  \,, 
  \label{eq:ggwwzz} 
\end{eqnarray} 
where the functions $\xi^{(n)}_{\hat{V},i}$  
($V=W,Z; i=1,2$) are obtained from the following expressions  
\begin{eqnarray} 
  \xi^{(n)}_{\hat W,1} + \xi^{(n)}_{\hat W, 2} =   
  \frac{M_W}{2}   \left. \partial^2_{p} \partial_{p_\rho}  
  \g^{(n)}_{\hat{G}^+ \hat{W}^-_\rho }(p) \right|_{p=0}  
  \,,~~~~~~~
  \xi^{(n)}_{\hat Z,1} + \xi^{(n)}_{\hat Z, 2} =   
  i \frac{M_Z}{2}   \left. \partial^2_{p} \partial_{p_\rho}  
  \g^{(n)}_{\hat{G}^0 \hat{Z}_\rho }(p) \right|_{p=0}  
  \,. 
\label{eq:bfm_2.1} 
\end{eqnarray} 
Equation ~(\ref{eq:bfm_2.1}) only fixes the sum of the coefficients 
$\xi^{(n)}_{\hat{V}, 1}$ and $\xi^{(n)}_{\hat{V}, 2}$. 
Note, however, that one still has the freedom  
of implementing suitable renormalization conditions 
for the two-point functions. For example, one can decide to  
renormalize the $W$- or the $Z$-background two-point functions on-shell.  
This fixes the difference of  
$\xi^{(i)}_{\hat{V}, 1}$ and 
$\xi^{(i)}_{\hat{V}, 2}$.  
However, as is well known~\cite{bkg,msbkg,grassi},  
the wave function renormalization of the background fields  
is related to the coupling constant renormalization and, therefore, the  
WTIs for three-point functions, discussed in the next section, provide  
a proper renormalization of the difference. 
 
The coefficients $\xi_{M_V}^{(n)}$ ($V=W/Z$) are 
fixed via the normalization conditions. In the case of the  
on-shell scheme~\cite{Sirlin:1980nh,sirlin-masses}, 
where the pole mass\footnote{As was proven in~\cite{grassi} this is 
  equivalent to the pole renormalization for the quantum gauge 
  bosons.}  
enters as a parameter, they would look as follows ($i=1,2$) 
\begin{eqnarray} 
  \gg^{(n),T}_{\hat W^+ \hat W^-}(p^*) =0 
  &~~~~{\rm with}~~~&  
  {\rm Re}(p^*)^2 = M_W^2\,\, 
  \nonumber\\  
  {\rm det}  
  \left( \begin{array}{cc}  
      \gg^{(n),T}_{\hat Z \hat Z}(p^*) &   \gg^{(n),T}_{\hat Z \hat A}(p^*) 
      \vspace{.2cm} \\   
      \gg^{(n),T}_{\hat A \hat Z}(p^*) &   \gg^{(n),T}_{\hat A \hat A}(p^*)  
    \end{array} \right)  =0 
  &~~~~{\rm with}~~~& 
  {\rm Re}(p^*)^2 = M_Z^2 
  \,, 
  \label{nor_con_W.0} 
\end{eqnarray} 
where the superscript $T$ marks the transversal parts. 
In this framework $M_W^2$ and $M_Z^2$ are the physical masses, which  
also serve as input parameters.  
 
Besides mass renormalizations, we have to 
take into account the renormalization of the photon self-energy and its mixing 
with the $Z$ bosons. The structure of  
counterterms for the mixed two-point functions and for the photon 
two-point function   
can be found in the literature \cite{BoeHolSpi86,krau_ew,grassi}. 
We can impose the following normalization conditions  
\begin{eqnarray} 
  \label{nor_con_W.00} 
  \gg^{(n),T}_{\hat A \hat A}(0) =0, ~~~~  \gg^{(n),T}_{\hat A \hat Z}(0) =0\,. 
\end{eqnarray} 
 
At the end of this section we want to note that at one- and two-loop order 
the Green functions 
$\g^{(n)}_{\hat W^+_\mu \hat W^-_\nu}(p)$ and 
$\g^{(n)}_{\hat Z_\mu \hat Z_\nu}(p)$  
in Eq.~(\ref{eq:ggwwzz}) could be chosen to be 
already symmetric, since, in the case of the two-point  
functions, there exists an effectively invariant regularization.  
It can be shown that in this case the naive prescription 
of $\gamma_5$ accompanied with dimensional regularization leads  
to the correct answer (see, e.g., \cite{gambino}). 
The symmetry is destroyed by Taylor subtraction and again 
restored by the counterterms which means that the quantities 
in~(\ref{eq:ggwwzz}) can be written as 
\begin{equation} 
  \gg_{\hat{V}_1 \hat{V}_2} (p) \,\,=\,\, (1-T^2_p)  
  \gg_{\hat{V}_1 \hat{V}_2} (p) + \Xi_{\hat{V}_1 \hat{V}_2} 
  \,. 
  \label{symtwo} 
\end{equation}  
Thus, in this case the method of algebraic 
renormalization is not needed. However,  
the calculation of the counterterms $\Xi_{\hat{V}_1 \hat{V}_2}$  
in Eq.~(\ref{symtwo})  
is necessary as the renormalization of the three-point  
functions with the help of Taylor subtraction --- which we present in the 
next section --- depends on these counterterms. 
 
 
\section{\label{sec:charged}Charged gauge boson vertices} 
 
\subsection{The case AWW} 
\label{AWW} 
 
The analysis of the vertex AWW turns out to be the most important for 
phenomenological studies, and it entails several interesting 
features.  In particular, in contrast to the previous examples, AZZ and ZZZ, 
the WTIs for the AWW amplitudes appear more cumbersome because of the  
presence of two-point functions.  
In the conventional algebraic renormalization methods this is very 
cumbersome, as all Green functions appearing in the WTIs have to be 
computed --- often for off-shell momenta. We will see that in our 
approach only a few Green functions remain which have to be evaluated 
with zero external momentum.  
 
In order to obtain a closed set of identities we have to differentiate 
w.r.t. $\lambda_A \hat{W}^+_\rho \hat{W}^-_\sigma$ and  
$\hat{A}_\mu \lambda_+\hat{W}^-_\sigma$.  
The first option leads to 
\begin{eqnarray} 
  i \left(p_{+} + p_{-}\right)^\mu \g^{(n)}_{\hat{A}_{\mu} \hat 
  W^{+}_{\rho}   
    \hat W^{-}_{\sigma}} (p_{+},p_{-})  
  \hphantom{xxxxxxxxxxxxxxxxxxxx} 
  \nonumber\\\mbox{} 
  - i e \left(  \gg^{(n)}_{\hat 
  W^{+}_{\rho}   
      \hat  W^{-}_{\sigma}} (p_{-}) - \gg^{(n)}_{\hat W^{+}_{\rho} \hat 
  W^{-}_{\sigma}} (-p_{+})   
  \right)  
  &=& \Delta^{\prime,(n)}_{W,\lambda_A \hat W^{+}_{\rho} \hat W^{-}_{\sigma}} 
  (p_{+},p_{-})  
  \,,  \nonumber \\ 
  i \left( p_{+} + p_{-} \right)^\mu \g^{(n)}_{\hat{A}_{\mu} \hat G^{+}   
    \hat  G^{-}} (p_{+},p_{-})  
  \hphantom{xxxxxxxxxxxxxxxxxxxx} 
  \nonumber\\\mbox{} 
  - i e \left( \gg^{(n)}_{\hat G^{+} \hat 
  G^{-}}   
    (p_{-}) - \gg^{(n)}_{\hat G^{+} \hat G^{-}}(-p_{+})   
  \right) &=& \Delta^{\prime,(n)}_{W,\lambda_A  \hat G^{+} \hat  G^{-}} 
  (p_{+},p_{-})   
  \,,\nonumber\\  
  i \left( p_{+} + p_{-} \right)^\mu \g^{(n)}_{\hat{A}_{\mu}   
    \hat  W^{+}_{\rho}  \hat G^{-} } (p_{+},p_{-})   
  \hphantom{xxxxxxxxxxxxxxxxxxxx} 
  \nonumber\\\mbox{} 
  - i e \left( 
  \gg^{(n)}_{  
      \hat  W^{+}_{\rho} \hat G^{-}} (p_{-}) - \gg^{(n)}_{\hat 
  W^{+}_{\rho} \hat G^{-}} (-p_{+})   
  \right) &=& \Delta^{\prime,(n)}_{W,\lambda_A  \hat  W^{+}_{\rho}  \hat G^{-} 
    }  
  (p_{+},p_{-})  
  \,. 
  \label{wti:aGG} 
\end{eqnarray}  
In analogy we obtain, from the functional differentiation w.r.t. 
$\hat{A}_\mu \lambda_+ \hat{W}^-_\sigma$: 
\begin{eqnarray} 
  -i p_{+}^\rho \g^{(n)}_{\hat{A}_{\mu} \hat W^{+}_{\rho}  
    \hat W^{-}_{\sigma}} (p_{+},p_{-}) + i M_W  
  \g^{(n)}_{\hat{A}_{\mu} \hat G^{+}  
    \hat W^{-}_{\sigma}} (p_{+},p_{-})  
  \hphantom{xxxxxxxxxxx} 
  \nonumber\\\mbox{} 
  + i e \left(  \gg^{(n)}_{\hat W^{+}_{\mu}  
      \hat  W^{-}_{\sigma}} (p_{-})  
  - \gg^{(n)}_{\hat A_{\mu} \hat 
      A_{\sigma}} (-p_{+}-p_{-})   
    + {c_W \over s_W} \gg^{(n)}_{\hat A_{\mu} \hat Z_{\sigma}} 
    (-p_{+}-p_{-})   
  \right)  
  &=& \Delta^{\prime,(n)}_{W,A_\mu \lambda_{+} \hat W^{-}_{\sigma}}  
  (p_{+},p_{-})  
  \,, \nonumber\\ 
  -i p_{+}^\rho \g^{(n)}_{\hat{A}_{\mu} \hat W^{+}_{\rho}  
    \hat G^{-}} (p_{+},p_{-}) + i M_W  
  \g^{(n)}_{\hat{A}_{\mu} \hat G^{+}  
    \hat G^{-}} (p_{+},p_{-})  
  \hphantom{xxxxxxxxxxxx} 
  \nonumber\\\mbox{}  
 + e\left( i \gg^{(n)}_{\hat W^+_{\mu} \hat G^-} (p_{-})  
 + {1\over 2 s_W} \gg^{(n)}_{\hat A_{\mu} \hat G^0} (-p_{+}-p_{-}) 
  \hphantom{xxxxxxxxxxxx} 
  \right.\nonumber\\\left.\mbox{}  
 - {i \over 2 s_W} \gg^{(n)}_{\hat A_{\mu} \hat H} (-p_{+}-p_{-}) 
        \right)          
  &=&  
  \Delta^{\prime,(n)}_{W,A_\mu \lambda_{+} \hat G^{-}} (p_{+},p_{-}) 
  \,. 
  \nonumber\\ 
  \label{wti:AwW}    
\end{eqnarray}  
Again all equations that  can be obtained by interchanging $W^+$ and $W^-$ 
are omitted. 
In the above equations it is assumed that the two-point functions  
are already correctly renormalized according to (\ref{eq:ggwwzz}). 
Therefore a prime is added to the corresponding $\Delta$ on the r.h.s.. 
 
By performing a Taylor subtraction $(1-T^2_{p^+,p^-})$, the above equations 
lead to the following universal breaking terms, written   
in terms of the Taylor-subtracted Green functions $\gh^{(n)}$:  
\begin{eqnarray} 
  \Psi^{\prime,(n)}_{\lambda_A \hat W^{+}_{\rho} \hat W^{-}_{\sigma}} 
  (p_{+},p_{-})  
  &=&   
  - i e 
  \left(  
    T^2_{p^-} \gg^{(n)}_{\hat W^{+}_{\rho} \hat  W^{-}_{\sigma}} (p_{-}) - 
    T^2_{p^+} \gg^{(n)}_{\hat W^{+}_{\rho} \hat W^{-}_{\sigma}} (-p_{+})  
  \right) 
  \nonumber\\&=& 
   i e 
  \left[ \left( \xi^{(n)}_{\hat{W},1} \, p^2_-g_{\rho\sigma}  +  
      \xi^{(n)}_{\hat{W},2} \, p_{-,\rho} p_{-,\sigma} \right)  
    -  \left( \xi^{(n)}_{\hat{W},1}  \, p^2_+g_{\rho\sigma} + 
  \xi^{(n)}_{\hat{W},2} \, p_{+,\rho} p_{+,\sigma}  
  \right) \right] 
  \,,      
  \nonumber\\ 
  \Psi^{\prime,(n)}_{\lambda_A  \hat G^{+} \hat  G^{-}} (p_{+},p_{-}) &=& 
  \Psi^{\prime,(n)}_{\lambda_A  \hat  W^{+}_{\rho}  \hat G^{-} } 
  (p_{+},p_{-})    
  \,\,=\,\,0\,, 
  \nonumber\\ 
  \Psi^{\prime,(n)}_{\hat A_\mu \lambda_{+} \hat W^{-}_{\sigma}} 
(p_{+},p_{-}) &=& i M_W 
\left( T^2_{p^+,p^-} - T^1_{p^+,p^-} \right)   
  \g^{(n)}_{\hat{A}_{\mu} \hat G^{+}  
    \hat W^{-}_{\sigma}} (p_{+},p_{-})  + i e  T^2_{p^-} \gg^{(n)}_{\hat 
    W^{+}_{\mu} \hat  W^{-}_{\sigma}} (p_{-}) 
 \nonumber \\ &=&  
{i\, M_W \over 2}  \left(  
    p_+^\rho p_+^\nu \partial_{p_+^\rho}  \partial_{p_+^\nu} +  
    2 \, p_+^\rho p_-^\nu \partial_{p_+^\rho}  \partial_{p_-^\nu}  
    + p_-^\rho p_-^\nu \partial_{p_-^\rho}  \partial_{p_-^\nu}  
  \right)       
  \nonumber\\&&\mbox{} 
  \left.  \g^{(n)}_{\hat{A}_{\mu} \hat G^{+}  
      \hat W^{-}_{\sigma}} (p_{+},p_{-}) \right|_{p_\pm=0}  
  - i e \left( \xi^{(n)}_{\hat{W}, 1} p^2_-g_{\mu\sigma} +  
  \xi^{(n)}_{\hat{W}, 2} p_{-,\mu} p_{-,\sigma} \right)  
  \,, 
  \nonumber \\ 
  \Psi^{\prime,(n)}_{\hat A_\mu \lambda_{+} \hat G^{-}} (p_{+},p_{-}) 
  &=& 0 \,.   
  \label{psi:aWW.1}  
\end{eqnarray}  
The r.h.s. of the last term in Eq.~(\ref{psi:aWW.1})  
vanishes because of covariance and zero-momentum subtraction. 
Notice the appearance of the breaking terms for the $W$ boson 
two-point functions. In principle also the corresponding ones from the 
charged Goldstone boson could appear. 
However, this can be avoided as  
zero momentum subtraction of the Goldstone self-energies automatically 
preserves the respective WTIs (cf. Ref.~\cite{GraHurSte99}). 
Note that due to the fact that the photon is massless 
there is no contribution from  
$\gg^{(n)}_{\hat A_{\mu} \hat H}$. In our  
approximation it furthermore doesn't contribute due to CP violation.

In order to restore the WTIs  
one has the freedom of adding  a non-invariant counterterm to the  
Green function $\g_{\hat A_\mu \hat W^+_\rho \hat W^-_\sigma}$. However, 
to remove   
$\Psi^{(n)}_{A_\mu \lambda_{+} \hat W^{-}_{\sigma}} (p_{+},p_{-})$ it 
is necessary   
that the latter is independent of  
$p^2_\pm \, g_{\mu\sigma}$ and $p_{\pm,\mu}p_{\pm,\sigma}$. This   
can be achieved by fixing the difference between the parameters 
$\xi^{(n)}_{\hat{W},1}-\xi^{(n)}_{\hat{W},2}$.   
Requiring that WTIs be preserved  
in the tree-level form amounts to imposing  a charge renormalization. 
 
In a next step we want to translate the information about the breaking 
terms into counterterms that will  restore the symmetry of the Green functions. 
In general the QAP allows for all possible breaking terms with 
suitable dimensions. However, not all of them are independent. 
The consistency conditions can be used to figure out the independent 
ones and thus reduce the counterterms to a minimal set. 
 
In the AWW case the most general counterterm that  can be used to 
re-absorb the  breaking term is of the form   
\begin{eqnarray} 
\label{c_aww} 
\Xi^{(n)}_{\hat A_\mu \hat W^+_\nu \hat W^-_\rho}(p^+,p^-) &=& 
e\left[ 
g_{\mu \nu} \left( \xi^{(n)}_{1} p^+_\rho +  \xi^{(n)}_{2} p^-_\rho \right)  +  
g_{\nu \rho} \left( \xi^{(n)}_{3} p^+_\mu +  \xi^{(n)}_{4} p^-_\mu 
\right)  
\right.\nonumber\\&&\left.\mbox{} 
 +  
g_{\mu \rho} \left( \xi^{(n)}_{5} p^+_\nu +  \xi^{(n)}_{6} p^-_\nu 
\right) 
\right]\,,  
\end{eqnarray} 
where the coefficients $\xi_i^{(n)}$ can be expressed through the  
breaking terms as we will show in the following. 
Owing to the first equations of~(\ref{wti:aGG}) and~(\ref{psi:aWW.1}) 
one obtains 
\begin{eqnarray} 
  \label{c_aww.1} 
  &&\xi^{(n)}_1 +  \xi^{(n)}_6=0\,,\,\,\, 
    \xi^{(n)}_2 +  \xi^{(n)}_5=0\,,\,\,\, 
    -\xi^{(n)}_4 = \xi^{(n)}_3 = \xi^{(n)}_{\hat W,1}\,,  
  \nonumber \\ 
  &&-(\xi^{(n)}_2 +  \xi^{(n)}_6) = \xi^{(n)}_1 +  \xi^{(n)}_5 
    = \xi^{(n)}_{\hat W,2}\,. 
\end{eqnarray} 
Note, that the sum of $\xi^{(n)}_{\hat W,1}+\xi^{(n)}_{\hat W,2}$ 
is given in Eq.~(\ref{eq:bfm_2.1}).  
The contraction of~(\ref{c_aww}) 
with $p_+^\rho$ and the comparison with  
$\Psi^{(n)}_{\hat A_\mu \lambda_{+} \hat W^{-}_{\sigma}}$ 
in~(\ref{psi:aWW.1}) leads to another set of equations. 
At first sight there are more equations than unknowns. 
However, one should have in mind that not all equations are 
independent due to WTIs like 
\begin{eqnarray} 
  \left. \partial_{p^-_\alpha}  \partial_{p^-_\alpha} 
  \g^{(n)}_{A_\mu G^+ W^-_\mu}(p^+,p^-) \right|_{p^\pm=0}  +  
  2 \,  
  \left. \partial_{p^-_\alpha}  \partial_{p^-_\beta} 
  \g^{(n)}_{A_\alpha G^+ W^-_\beta}(p^+,p^-) \right|_{p^\pm=0}  
  \nonumber\\ 
  - 
  e  \left. \partial_{p_\alpha}  \partial_{p_\alpha} \partial_{p_\mu} 
  \g^{(n)}_{G^+ W^-_\mu}(p) \right|_{p=0}  
  &=& 0\,, 
\end{eqnarray} 
which is a special case of 
\begin{eqnarray} 
\label{cond_cond.4} 
\left( \left. \partial_{p_\alpha} \partial_{p_\beta} \g^{(n)}_{\hat 
V^a_\gamma \hat \Phi_i   
\hat V^b_\nu}\right|_{p=q=0} +  
\left. \partial_{p_\beta} \partial_{p_\gamma} \g^{(n)}_{\hat 
V^a_\alpha \hat \Phi_i   
\hat V^b_\nu}\right|_{p=q=0} +  
\left. \partial_{p_\gamma} \partial_{p_\alpha} \g^{(n)}_{\hat 
V^a_\beta \hat \Phi_i   
\hat V^b_\nu}\right|_{p=q=0} 
\right) \nonumber \\ 
+ \sum_j M_{a,j} \left. \partial_{p_\alpha} \partial_{p_\beta} 
\partial_{p_\gamma}   
\g^{(n)}_{\hat \Phi_j \hat \Phi_i \hat V^b_\nu}(p,q) \right|_{p=q=0} +  
f^{abx} \left. \partial_{p_\alpha} \partial_{p_\beta} \partial_{p_\gamma}  
\g^{(n)}_{\hat \Phi_i \hat V^x_\nu}(-p) \right|_{p=0} &=& 0 \,. 
\end{eqnarray} 
The sum runs over all would-be-Goldstone fields  
$G^0$ and $G^\pm$ with masses $M_{\pm, G^\pm} = \pm i M_W$, 
$M_{Z, G^0} = - M_Z$  
and zero for all the other combinations. $V^a_\alpha$ denote the  
gauge fields, where $\alpha$ runs over the index of the adjoint  
representation for $SU(3)\times SU(2) \times U(1)$.   
Here, $f^{abx}$ represent the structure constants of the 
gauge group in the adjoint representation. 
Other useful identities can be easily obtained by 
differentiating with respect to $p_\alpha,p_\beta,q_\gamma$ or 
$p_\alpha,q_\beta,q_\gamma$ or $q_\alpha,q_\beta,q_\gamma$.  Notice 
that all Green functions appearing in Eq.~(\ref{cond_cond.4}) are 
superficially finite. 
 
Having this in mind we can write down the results for  
$\xi^{(n)}_i, (i=1,...6)$ 
\begin{eqnarray} 
  \label{c_aww.5} 
  ie\xi^{(n)}_1 &=& -ie\xi^{(n)}_6 =  
  {1\over 18}\left( - 6 {\cal M}_1 - {\cal M}_2 + 5 {\cal M}_3 \right) 
  \,, 
\nonumber\\
  ie\xi^{(n)}_3 &=& -ie\xi^{(n)}_4 = 
  {1\over 18}\left( 5 \, {\cal M}_2  - {\cal M}_3\right)  
  \,, 
\nonumber\\
  ie\xi^{(n)}_5 &=& -ie\xi^{(n)}_2 =  
  {1\over 18} \left( 6 {\cal M}_1 - {\cal M}_2 - \, {\cal M}_3 
  \right)  
  \,,  
\end{eqnarray} 
where we have defined 
\begin{eqnarray} 
  \label{c_aww.2} 
  {\cal M}_1 &=&    \frac{i M_W}{16}  
   \left. \partial_{p^+_\alpha}  \partial_{p^+_\alpha} 
   \g^{(n)}_{A_\mu G^+ W^-_\mu}(p^+,p^-) \right|_{p^\pm=0} 
  \,,\nonumber \\  
  {\cal M}_2 &=&   \frac{i M_W}{16}  
   \left. \partial_{p^-_\alpha}  \partial_{p^-_\alpha} 
   \g^{(n)}_{A_\mu G^+ W^-_\mu}(p^+,p^-) \right|_{p^\pm=0} 
  \,,\nonumber \\  
  {\cal M}_3 &=&   \frac{i M_W}{8}  
  \left. \partial_{p^-_\alpha}  \partial_{p^-_\beta} 
  \g^{(n)}_{A_\alpha G^+ W^-_\beta}(p^+,p^-) \right|_{p^\pm=0} 
  \,. 
\end{eqnarray} 
Note that it is also possible to find other representations 
of the results. However, they are equivalent after exploiting 
Eq.~(\ref{cond_cond.4}). 
 
Finally, the symmetrical Green function reads 
\begin{eqnarray} 
  \label{ct_AWW} 
  &&\gg^{(n)}_{\hat A_\rho \hat W^+_\mu \hat W^-_\nu}(p^+,p^-) = \nonumber \\  
  &&\mbox{}~~~ \g^{(n)}_{\hat A_\rho \hat W^+_\mu \hat W^-_\nu}(p^+,p^-)  
  - \Big[ T^2_{p^+,p^-} \g^{(n)}_{\hat A_\rho \hat W^+_\mu \hat 
  W^-_\nu}(p^+,p^-) + 
 \Xi^{(n)}_{\hat A_\rho \hat W^+_\mu \hat W^-_\nu}(p^+,p^-) \Big] 
  \,.   
\end{eqnarray}   
 
Let us summarize the steps which that  to be 
performed in order to compute the ${\cal O}(\alpha\alpha_s)$ corrections to 
the $AWW$ vertex. The basic equation is Eq.~(\ref{eq:gg}). 
In a first step the two-loop amplitude, denoted  by 
$\Gamma^{(2)}_{AWW}$, has to be   
calculated using a specific regularization scheme and a specific subtraction  
scheme. There are two contributions: genuine two-loop diagrams and 
finite one-loop counterterm diagrams.  
Clearly, the same regularization scheme has to be used 
in the two   
contributions. 
The divergent parts  in both contributions are assumed to be subtracted  
already. The finite one-loop counterterm contributions  
are two-fold: first, the one-loop counterterm due to the Taylor-subtraction  
and second, the universal one-loop counterterms  
(see Eqs.~(\ref{one_5})--(\ref{one_6})). 
 
 
\subsection{\label{ZWW}The case ZWW} 
 
Having the physical process $e^+e^- \rightarrow W W$ in mind,  
we also have to discuss the QCD corrections to the ZWW vertex 
within the channel $e^+e^- \rightarrow Z \rightarrow W W$.  
The case ZWW has some similarity to AWW. However, due to the 
connection of the $Z$ boson and the neutral Goldstone boson 
with the finite $Z$ boson mass, both the identities and the analysis, 
get more involved. 
 
The equations corresponding to~(\ref{wti:aGG}) and~(\ref{wti:AwW}) read: 
\begin{eqnarray} 
  i \left(p_{+} + p_{-}\right)^\mu \g^{(n)}_{\hat{Z}_{\mu} \hat W^{+}_{\rho}  
   \hat W^{-}_{\sigma}} (p_{+},p_{-})  
  - M_Z  \g^{(n)}_{\hat{G}^{0} \hat W^{+}_{\rho}  
  \hat W^{-}_{\sigma}} (p_{+},p_{-})  
  \nonumber\\ 
  + ie {c_W \over s_W} \left(  \gg^{(n)}_{\hat W^{+}_{\rho}  
  \hat  W^{-}_{\sigma}} (p_{-}) - \gg^{(n)}_{\hat W^{+}_{\rho} \hat 
   W^{-}_{\sigma}} (-p_{+})   
  \right)  
  &=&  
  \Delta^{\prime,(n)}_{W,\lambda_Z \hat W^{+}_{\rho} \hat W^{-}_{\sigma}} 
   (p_{+},p_{-})  
  \,, 
\nonumber  \\ 
  i \left( p_{+} + p_{-} \right)^\mu \g^{(n)}_{\hat{Z}_{\mu} \hat G^{+}  
  \hat  G^{-}} (p_{+},p_{-})  
  - M_Z   \g^{(n)}_{\hat{G}^{0} \hat G^{+}  
  \hat  G^{-}} (p_{+},p_{-})  
  \nonumber\\ 
  +i e { c^2_W - s^2_W\over 2\, s_W c_W} \left( \gg^{(n)}_{\hat G^{+} \hat G^{-}}  
  (p_{-}) - \gg^{(n)}_{\hat G^{+} \hat G^{-}}(-p_{+})  
  \right)  
  &=&  
  \Delta^{\prime,(n)}_{W,\lambda_Z  \hat G^{+} \hat  G^{-}} (p_{+},p_{-}) 
  \,, 
\nonumber  \\ 
  \label{wti:zGW} 
  i \left( p_{+} + p_{-} \right)^\mu \g^{(n)}_{\hat{Z}_{\mu}  
  \hat  W^{+}_{\rho}  \hat G^{-} } (p_{+},p_{-})   
  - M_Z   \g^{(n)}_{\hat{G}^{0}  
  \hat  W^{+}_{\rho}  \hat G^{-} } (p_{+},p_{-})    
  \nonumber\\ 
  +i e   \left( {c_W \over s_W} \gg^{(n)}_{ 
  \hat  W^{+}_{\rho} \hat G^{-}} (p_{-}) -  
 { c^2_W - s^2_W\over 2\, s_W c_W} 
\gg^{(n)}_{\hat 
  W^{+}_{\rho} \hat G^{-}} (-p_{+})   
  \right)  
  &=& 
  \Delta^{\prime,(n)}_{W,\lambda_A  \hat  W^{+}_{\rho}  \hat G^{-} } (p_{+},p_{-}) 
  \,, 
\end{eqnarray}  
and  
\begin{eqnarray} 
  -i p_{+}^\rho \g^{(n)}_{\hat{Z}_{\mu} \hat W^{+}_{\rho}  
    \hat W^{-}_{\sigma}} (p_{+},p_{-}) + i M_W  
  \g^{(n)}_{\hat{Z}_{\mu} \hat G^{+}  
    \hat W^{-}_{\sigma}} (p_{+},p_{-})  
  \hphantom{xxxxxxxxxxx} 
  \nonumber\\\mbox{} 
  - i e \left({c_W \over s_W}  \gg^{(n)}_{\hat W^{+}_{\mu}  
      \hat  W^{-}_{\sigma}} (p_{-})  
  + \gg^{(n)}_{\hat Z_{\mu} \hat 
      A_{\sigma}} (-p_{+}-p_{-})   
  \hphantom{xxxxxxxxxxx} 
  \right.\nonumber\\\left.\mbox{} 
    - {c_W \over s_W} 
    \gg^{(n)}_{\hat Z_{\mu} \hat Z_{\sigma}} 
    (-p_{+}-p_{-})   
  \right)  
  &=& \Delta^{\prime,(n)}_{W,Z_\mu \lambda_{+} \hat W^{-}_{\sigma}}  
  (p_{+},p_{-})  
  \,, \nonumber\\ 
  -i p_{+}^\rho \g^{(n)}_{\hat{Z}_{\mu} \hat W^{+}_{\rho}  
    \hat G^{-}} (p_{+},p_{-}) + i M_W  
  \g^{(n)}_{\hat{Z}_{\mu} \hat G^{+}  
    \hat G^{-}} (p_{+},p_{-})  
  \hphantom{xxxxxxxxxxxx} 
  \nonumber\\\mbox{}  
+ e\left(- i{c_W \over s_W}  \gg^{(n)}_{\hat W^+_{\mu} \hat G^-} (p_{-})  
+ {1\over 2 s_W} \gg^{(n)}_{\hat Z_{\mu} \hat G^0} (-p_{+}-p_{-}) 
  \hphantom{xxxxxxxxxxxx} 
  \right.\nonumber\\\left.\mbox{}  
- {i\over 2 s_W} \gg^{(n)}_{\hat Z_{\mu} \hat H} (-p_{+}-p_{-}) 
        \right)          
  &=&  
  \Delta^{\prime,(n)}_{W,Z_\mu \lambda_{+} \hat G^{-}} (p_{+},p_{-}) 
  \,. 
  \nonumber\\ 
  \label{wti:ZwW}    
\end{eqnarray}  
Again all equations that can be obtained by interchanging $W^+$ and $W^-$ 
are omitted.  
 
In order to keep the discussion simpler we restrict ourselves to the one- and 
two-loop level. This means in the following equations the index $n$ is 
either 1 or 2. 
The second-order Taylor subtraction leads to  
\begin{eqnarray} 
  \label{psi:aWW.1_2}  
  \Psi^{\prime,(n),W}_{\lambda_Z \hat W^{+}_{\rho} \hat W^{-}_{\sigma}} 
  (p_{+},p_{-}) &=&   
  -{ M_Z \over 2}  \left(  
  p_+^\mu p_+^\nu \partial_{p_+^\mu}  \partial_{p_+^\nu} +  
  2 \, p_+^\mu p_-^\nu \partial_{p_+^\mu}  \partial_{p_-^\nu} +  
  p_-^\mu p_-^\nu \partial_{p_-^\mu}  \partial_{p_-^\nu}  
  \right)   
  \nonumber \\  
  &&  
  \left.  \g^{(n)}_{\hat{G}^{0} \hat W^{+}_{\rho}  
  \hat W^{-}_{\sigma}} (p_{+},p_{-}) \right|_{p_\pm=0}  
  - {i e \, c_W \over s_W}  
 \left[ \left( \xi^{(n)}_{\hat{W},1} \, p^2_-g_{\rho\sigma}  +  
      \xi^{(2)}_{\hat{W},2} \, p_{-,\rho} p_{-,\sigma} \right)  
 \right. \nonumber \\  
  && \left. 
    -  \left( \xi^{(n)}_{\hat{W},1}  \, p^2_+g_{\rho\sigma} + 
  \xi^{(n)}_{\hat{W},2} \, p_{+,\rho} p_{+,\sigma}  
  \right) \right]\,, 
  \nonumber \\ 
  \Psi^{\prime,(n),W}_{\lambda_Z \hat G^{+} \hat G^{-}} (p_{+},p_{-}) 
 &=& 0\,, 
  \nonumber \\ 
  \Psi^{\prime,(n),W}_{\lambda_Z \hat W^{+}_{\rho} \hat G^{-}} (p_{+},p_{-}) 
&=& 0\,,  
\nonumber \\ 
  \Psi^{\prime,(n)}_{\hat Z_\mu \lambda_{+} \hat W^{-}_{\sigma}} 
 (p_{+},p_{-})  
&=& {i\, M_W \over 2}  \left(  
    p_+^\rho p_+^\nu \partial_{p_+^\rho}  \partial_{p_+^\nu} +  
    2 \, p_+^\rho p_-^\nu \partial_{p_+^\rho}  \partial_{p_-^\nu}  
    + p_-^\rho p_-^\nu \partial_{p_-^\rho}  \partial_{p_-^\nu}  
  \right)       
  \nonumber\\&&\mbox{} 
  \left.  \g^{(n)}_{\hat{Z}_{\mu} \hat G^{+}  
      \hat W^{-}_{\sigma}} (p_{+},p_{-}) \right|_{p_\pm=0}    
  + i e \frac{c_W}{s_W}\left( \xi^{(n)}_{\hat{W}, 1} p^2_-g_{\mu\sigma} +  
  \xi^{(n)}_{\hat{W}, 2} p_{-,\mu} p_{-,\sigma} \right)   
\nonumber\\&&\mbox{} 
  - i e {c_W \over s_W} \left( \xi^{(n)}_{\hat{Z}, 1} (p_- + p_+)^2 \, 
  g_{\mu\sigma} +   
  \xi^{(n)}_{\hat{Z}, 2} (p_- + p_+)_{\mu} (p_- + p_+)_{\sigma} \right)  
  \,, 
  \nonumber \\ 
  \Psi^{\prime,(n)}_{Z_\mu \lambda_{+} \hat G^{-}} (p_{+},p_{-}) &=& 0 \,. 
\end{eqnarray}  
In general the breaking term 
$\Psi^{(n),W}_{\lambda_Z \hat W^{+}_{\rho} \hat G^{-}}$  
does not vanish. However, as we only consider two-loop  
QCD corrections it is zero. 
In this approximation also the Green function  
$\g^{(n)}_{\hat{G}^{0} \hat G^{+} \hat G^{-}} (p_{+},p_{-})$  
(and thus $\Psi^{\prime,(n),W}_{\lambda_Z \hat G^{+} \hat G^{-}} 
(p_{+},p_{-})$)  
vanishes as we checked by an explicit 
calculation. This is essentially due to the invariance under CP 
transformations of the bosonic sector. 
Note that starting form three loops, the CP violation  
induced by the CKM mixings will generate some CP  
violating bosonic counterterms.  
 
The most general counterterm that can be used to 
re-absorb  the  breaking term of the WTIs reads 
\begin{eqnarray} 
\label{c_zww} 
\Xi^{(n)}_{\hat Z_\mu \hat W^+_\rho \hat W^-_\sigma}(p^+,p^-) &=&  
e\left[ 
g_{\mu \rho} \left( \xi^{(n)}_{7} p^+_\sigma +  \xi^{(n)}_{8} 
p^-_\sigma \right)  +   
g_{\rho \sigma} \left( \xi^{(n)}_{9} p^+_\mu +  \xi^{(n)}_{10} p^-_\mu \right)  
\right.\nonumber\\&&\left.\mbox{} 
+ g_{\mu \sigma} \left( \xi^{(n)}_{11} p^+_\rho +  \xi^{(n)}_{12} p^-_\rho 
  \right) 
\right] 
\,. 
\end{eqnarray} 
 
Inserting $\Xi^{(n)}_{\hat Z_\mu \hat W^+_\nu \hat W^-_\rho}(p^+,p^-)$ 
in the above identities and comparing the coefficients with the 
breaking terms leads to 
equations from which the coefficients $\xi^{(n)}_i (i=7,\dots,12)$ 
can be determined. 
One possible set of equations reads 
\begin{eqnarray} 
  ie\left(4 \xi^{(n)}_9 +  \xi^{(n)}_7 + \xi^{(n)}_{11}\right)  
  &=& - {1 \over 8}  
  \partial_{p^+_\alpha}\partial_{p^+_\alpha} {\cal M}_{\beta\beta}  \,, 
  \nonumber \\ 
  ie\left(5 \xi^{(n)}_7 +  5 \xi^{(n)}_{11} + 2 \xi^{(n)}_9\right) 
  &=& - {1\over 4} 
  \partial_{p^+_\alpha}\partial_{p^+_\beta} {\cal M}_{\alpha\beta} \,, 
  \nonumber \\ 
  ie\left(5 \xi^{(n)}_7 +  5 \xi^{(n)}_9 + 2 \xi^{(n)}_{11}\right) 
  &=& {1\over 4} 
  \partial_{p^+_\alpha}\partial_{p^+_\beta} {\cal N}_{\alpha\beta} \,, 
  \nonumber\\ 
  ie\xi_8 &=& \frac{1}{72} 
    \left(5\partial_{p^-_\alpha}\partial_{p^+_\beta} 
          -\partial_{p^+_\alpha}\partial_{p^-_\beta} 
          -\partial_{p^+_\gamma}\partial_{p^-_\gamma}g^{\alpha\beta} 
    \right) {\cal M}_{\alpha\beta} 
  \,, 
  \nonumber\\ 
  ie\xi_{10} &=& \frac{1}{72} 
    \left(5\partial_{p^+_\alpha}\partial_{p^-_\beta} 
          -\partial_{p^-_\alpha}\partial_{p^+_\beta} 
          -\partial_{p^-_\gamma}\partial_{p^+_\gamma}g^{\alpha\beta} 
    \right) {\cal M}_{\alpha\beta} 
  \,, 
  \nonumber\\ 
  ie\xi_{12} &=& \frac{1}{72} 
    \left(5\partial_{p^+_\gamma}\partial_{p^-_\gamma}g^{\alpha\beta} 
          -\partial_{p^-_\alpha}\partial_{p^+_\beta} 
          -\partial_{p^+_\alpha}\partial_{p^-_\beta} 
    \right) {\cal M}_{\alpha\beta} 
  \,, 
\label{eq:zww_1} 
\end{eqnarray}  
where we introduced the short-hand notation 
${\cal M}_{\alpha\beta} =  
\Psi^{(n),W}_{\lambda_Z \hat W^{+}_{\alpha} \hat W^{-}_{\beta}} 
  (p_{+},p_{-})$ and  ${\cal N}_{\alpha\beta} =\Psi^{(n)}_{\hat 
Z_\alpha \lambda_{+} \hat W^{-}_{\beta}} (p_{+},p_{-})$. 
It is understood, that after the differentiation the momenta are set 
to zero. 
Note that the equations~(\ref{eq:zww_1}) 
are only unique up the consistency conditions (similar to the one in 
Eq.~(\ref{cond_cond.4})) among the WTIs.  
 
As in the previous case, the wave-function renormalization of 
the background field $\hat Z$ is fixed by the WTIs. In particular, in 
our analysis the $\theta_W$-angle is fixed by the on-shell condition  
$M_W / M_Z = c_W$, where $M_W$ and $M_Z$ are the physical masses.  
Actually, using the above equations one obtains two equations which 
fix $\xi^{(n)}_{\hat Z,i} (i=1,2)$ 
\begin{eqnarray} 
  \xi_{\hat{Z},1} &=& \xi_{\hat{W},1} + 
    \frac{s_W M_W}{36 e c_W} 
    \left( 
       -\frac{1}{4} \partial_{p^-_\mu} \partial_{p^-_\sigma} 
       +\frac{5}{8} \partial_{p^-_\alpha} \partial_{p^-_\alpha}g^{\mu\sigma} 
    \right)  
    \Gamma_{\hat{Z}_\mu\hat{G}^+\hat{W}_\sigma^-}(p_+,p_-)\Bigg|_{p_\pm=0} 
    \,, 
    \nonumber\\ 
  \xi_{\hat{Z},2} &=& \xi_{\hat{W},2} + 
    \frac{s_W M_W}{36 e c_W} 
    \left( 
                    \partial_{p^-_\mu} \partial_{p^-_\sigma} 
       -\frac{1}{4} \partial_{p^-_\alpha} \partial_{p^-_\alpha}g^{\mu\sigma} 
    \right)  
    \Gamma_{\hat{Z}_\mu\hat{G}^+\hat{W}_\sigma^-}(p_+,p_-)\Bigg|_{p_\pm=0} 
    \,, 
  \label{eq:bfm_2.2} 
\end{eqnarray} 
where $\xi^{(n)}_{\hat W,i}$ is given in Eq.~(\ref{c_aww.1}). 
Again, using the consistency conditions, one can check that 
other possible equations are not independent.  
 
Finally we want to remark that at higher orders also  
other counterterms may be involved in the analysis. 
However, this 
depends on the specific type of radiative corrections which are taken  
into account.  
In particular, $\g^{(3)}_{\hat G^0 \hat G^+\hat G^-}$ 
is only needed if three-loop electro-weak corrections are considered. 
 
 
\section{\label{IRR}IR re-shuffling} 
 
In practical applications quark masses can often be neglected. 
However, through the Taylor subtraction  
this can in general induce IR divergences in the  
corresponding two- and three-point functions. 
On the other hand it is important that the breaking terms are IR-finite. 
In this section we discuss the modifications of our procedure 
needed to deal with these cases. 
In a first step we want to approach the problem from  
a more theoretical point of view and only then 
apply it to the case of AWW. 
 
A careful analysis of the off-shell IR problems 
in the SM~\cite{krau_ew,grassi} 
shows that suitable normalization conditions are 
sufficient to guarantee the IR finiteness of Green functions 
in case non-exceptional momenta are chosen.  
However, Taylor subtraction around zero external momenta 
may cause problems.  
Let us assume that only the highest order of the derivative 
leads to IR divergences. 
This means that for a given Green function 
$\g$, with UV divergence degree $\omega$, 
$T^\omega \g$ is IR-divergent but 
$T^{\omega-1} \g$ is not. 
It is then tempting not to perform  
the complete Taylor expansion 
and leave out the term with highest power of derivative. 
This modifies the 
subtraction scheme presented in \cite{fg,GraHurSte99} 
as discussed in the following. 
 
Acting on a  broken WTI such as  Eq.~(\ref{WTI})  
with the Taylor operator $(1-T^\delta)$, we obtain 
\begin{equation} 
  \label{WTI.1} 
  (1-T^\delta)  {\cal W}_{(\lambda)} \left( \g^{(n)} \right)  
  \,\,=\,\,0 
  \,. 
\end{equation}  
After commuting the Taylor operator  
$(1-T^\delta)$ with ${\cal W}_{(\lambda)}$ 
this transforms to 
\begin{equation} 
  \label{WTI.2} 
  {\cal W}_{(\lambda)} \left[(1-T^{\delta^\prime} ) \g^{(n)} \right]   
  \,\,=\,\, 
  \left[T^{\delta}{\cal W}_{(\lambda)} - {\cal 
  W}_{(\lambda)}T^{\delta^\prime} \right]   
  \,\,\equiv\,\,  
  \hbar^n \Psi_W^{(n)}(\lambda)  
  \,, 
\end{equation}  
where $\delta^\prime$ is the naive power counting degree 
of $\g^{(n)}$. In general, one has $\delta  \geq   \delta^\prime$, 
Therefore the commutation of the Taylor operator with  
${\cal W}_{(\lambda)}$ leads to over-subtractions of  
$\g^{(n)}$ and, thus, to the new breaking terms $\Psi_W^{(n)}(\lambda)$  
occurring on the r.h.s. of Eq. (\ref{WTI.2}) (for more details see 
\cite{GraHurSte99}). 
 
In Eq. (\ref{WTI.2})  both terms, $T^{\delta^\prime} \g^{(n)}$  
and $\Psi_W^{(n)}(\lambda)$,  could be IR divergent.  
This suggests a re-shuffling of the terms from the  
r.h.s. to the l.h.s.,  which defines a new breaking term through 
\begin{equation} 
  \label{WTI.3} 
  {\cal W}_{(\lambda)} \left[(1-T^{\delta_{\rm IR}} ) \g^{(n)} \right] 
  \,\,=\,\,     
  \left[T^{\delta}{\cal W}_{(\lambda)} - {\cal 
  W}_{(\lambda)}T^{\delta^\prime} \right]   
  +  {\cal W}_{(\lambda)} \left[(T^{\delta^\prime} - T^{\delta_{\rm 
  IR}})  \g^{(n)} \right]  
  \,\,\equiv\,\,  
  \hbar^n \Psi_W^{\prime (n)}(\lambda)  
  \,. 
\end{equation}  
Note that $\delta_{IR} = \delta' -1$ is used for the  
IR-divergent Green functions. In 
this way all terms of Eq.~(\ref{WTI.3}) are IR-safe. 
The price of the 
re-shuffling is that the expression for $\Psi_W^{\prime (n)}(\lambda)$ 
becomes in general more complicated than  the original  
breaking term, $\Psi_W^{ (n)}(\lambda)$. 
However, its computation is still simpler than 
$\Delta_W^{(n)}(\lambda)$ of Eq.~(\ref{WTI}).  
The only requirement is 
an intermediate IR regulator, needed for the evaluation of  
the individual Green functions 
appearing in $\Psi_W^{\prime (n)}(\lambda)$. 
 
We also have to mention that the new breaking terms $\Psi_W^{\prime 
  (n)}(\lambda)$ and the corresponding counterterms $\Xi_W^{(n)}$ 
depend on the UV subtraction. In fact, since $\delta_{IR} = \delta' -1$, 
some Green functions are only superficially finite because of the UV 
subtraction. This implies that $\Psi_W^{\prime (n)}(\lambda)$  
as well as the final counterterm 
$-T^{\delta_{IR}}\g^{(n)} + \Xi^{\prime (n)}$  
(see~Eq.~(\ref{nor_con})) 
depends on the computation details.  
In addition, the dependence on the UV regulator of the   
breaking terms and  
the corresponding 
non-invariant counterterms breaks the universality  
of the computation (see~\cite{GraHurSte99}).  
 
As an explicit example let us consider the case AWW. 
In particular, we specify to the top-bottom doublet and neglect the 
mass of the bottom quark. 
In this case the one-loop sub-divergences (cf. Section.~\ref{one}) 
become IR-divergent. Using the IR re-shuffling discussed above 
Eq.~(\ref{one_1.1}) transforms to 
\begin{eqnarray} 
  \Psi'^{(1)}_{\lambda_A  \bar{b} b}(p,q)   &=&  
  i \left(p + q \right)^\mu T^0_{p,q} 
  \g^{(1)}_{\hat{A}_{\mu}  \bar{b} b}(p,q) 
  + i e Q_{q} \left( ( T^1_{q} -T^0_{q}) \g^{(1)}_{\bar{b} b}(q)   
    -  (T^1_{p}-T^0_{p})  
    \g^{(1)}_{\bar{b} b}(-p) \right)  
  \,, 
  \nonumber\\ 
  \Psi'^{(1)}_{\lambda_+  \bar{t} b}(p_t,p_b)   &=&  
  i\, M_W   
\left( T^1_{p_t p_b} - T^0_{p_t p_b} \right)   
    \g^{(1)}_{\hat G^{+} \bar{t} b}(p_t,p_b)  
  - { i e \over s_W\sqrt{2}} P_R V_{t b}  
    (T^1_{p_b}-T^0_{p_b}) 
   \g^{(1)}_{\bar{b}b}(p_b) 
  \,. 
  \label{psi:vqqp} 
\end{eqnarray} 
Of course, no IR divergences appear for the $A\bar{t}t$ vertex and thus we 
still have $\Psi^{(1)}_{\lambda_A  \bar{t} t}=0$. 
Notice that  
the advantages due to the zero-momentum subtractions have only slightly been 
reduced. The computation of the breaking terms  
still relies on Green functions expanded around zero external momenta. 
We also stress that the proposed rearrangement solves the spurious 
IR problem due to Taylor subtractions in general. 
 
The problem of IR divergences appears also at two loops. In that case, 
one has to handle the intermediate regularization procedure with some 
care.  For completeness concerning the AWW description, we present an 
example of IR re-shuffling at two loops. 
 
It is easy to see that the zero-momentum subtraction of the first 
of Eqs.~(\ref{wti:AwW}) and the breaking term 
$\Psi^{(2)}_{A_\mu \lambda_{+} \hat W^{-}_{\sigma}} (p_{+},p_{-})$ are 
IR-divergent in the approximation that $m_b=0$. So, we have to 
recombine the Green functions in such a way that the computation is  
IR-safe.  For that purpose, we can notice that also the zero-momentum 
subtraction of $\gg^{(2)}_{\hat A_{\mu} \hat Z_{\sigma}}$ and 
$\gg^{(2)}_{\hat A_{\mu} \hat A_{\sigma}}$ produces IR 
divergence. Therefore, the most natural IR-safe recombination is 
\begin{eqnarray} 
  \label{IR_1} 
  \Psi^{\prime, (2)}_{A_\mu \lambda_{+} \hat W^{-}_{\sigma}} 
  (p_{+},p_{-})  
  &=& 
  {i\, M_W \over 2}  \left(  
    p_+^\rho p_+^\nu \partial_{p_+^\rho}  \partial_{p_+^\mu} +  
    2 \, p_+^\rho p_-^\nu \partial_{p_+^\rho}  \partial_{p_-^\nu}  
   \right.\nonumber\\&&\left.\mbox{} 
     +p_-^\rho p_-^\nu \partial_{p_-^\rho}  \partial_{p_-^\nu}  
  \right)     
  \left.  \g^{(2)}_{\hat{A}_{\mu} \hat G^{+}  
      \hat W^{-}_{\sigma}} (p_{+},p_{-}) \right|_{p_\pm=0}  
  \nonumber\\&&\mbox{} 
  - i e 
  \left[ \left( \xi^{(n)}_{\hat{W},1} \, p^2_-g_{\rho\sigma}  +  
      \xi^{(n)}_{\hat{W},2} \, p_{-,\rho} p_{-,\sigma} \right)  
    -  \left( \xi^{(n)}_{\hat{W},1}  \, p^2_+g_{\rho\sigma} + 
  \xi^{(n)}_{\hat{W},2} \, p_{+,\rho} p_{+,\sigma}  
  \right) \right] 
  \nonumber\\&&\mbox{} - i e \left(  
  - T^2_{q} \gg^{(2)}_{\hat A_{\mu} \hat A_{\sigma}} (-q)   
    + {c_W \over s_W} T^2_{q} \gg^{(2)}_{\hat A_{\mu} \hat Z_{\sigma}}(-q)   
  \right) 
\,, 
\end{eqnarray} 
where $q= p_{+}+p_{-}$.  
$\Psi^{\prime, (2)}_{A_\mu \lambda_{+} \hat W^{-}_{\sigma}} $ is  
IR-safe  
and, thus, can be used to compute the counterterms. 
 
 
\section{Conclusion} 
 
The techniques developed in~\cite{GraHurSte99} have been applied to 
the three-gauge-boson vertices. In the framework of the BFM all 
functional identities are derived at the $n$-loop order.  
Since in the SM there exists no invariant regularization scheme  
(besides the lattice regularization with the Ginsparg-Wilson version 
of chiral symmetry) the functional identities  
are broken by local terms. 
Most of them are simply removed by the application of the Taylor  
operator~\cite{GraHurSte99}. The analysis of the remaining ones is 
presented up to the two-loop level,  where additional QCD corrections 
 to the one-loop fermionic diagrams are considered. 
Finally 
subtleties in connection to IR divergences resulting for the expansion 
around zero external momenta are discussed in detail. 
 
 
\section*{Acknowledgments} 
The research of P.A.G. is under the NSF grants no. PHY-9722083 and 
PHY-0070787. We thank A. Ferroglia and G. Ossola for useful  
discussions and a careful reading of the manuscript.  
 

\end{document}